\documentclass[american,aps,pra,amsmath,amssymb,showpacs, twocolumn]{revtex4-1}
\usepackage{colortbl}
\usepackage[T1]{fontenc}
\usepackage[utf8]{luainputenc}
\usepackage{amsmath}
\usepackage{amssymb}
\usepackage{xcolor}
\usepackage{graphicx}
\usepackage{verbatim}
\usepackage{setspace}

\pdfpageattr {/Group << /S /Transparency /I true /CS /DeviceRGB>>}

\def\be{\begin{equation}}
\def\ee{\end{equation}}
\def\re#1{(\ref{#1})}

\usepackage{mathbbol}
\usepackage{dsfont}

\makeatother

\usepackage{babel}

\global\long\def\ket#1{|#1\rangle}
\global\long\def\bra#1{\langle#1|}
\global\long\def\braket#1#2{\langle#1|#2\rangle}
\newcommand{\bracket}[2]{\langle #1 | #2 \rangle}
\global\long\def\ketbra#1#2{|#1\rangle\langle#2|}
\global\long\def\psiS{\psi_{S}}
\global\long\def\idmat{\mathds{1}}
\global\long\def\tr{\operatorname{tr}}
\global\long\def\sN{s_N}

\begin{document}

\title{Multiqubit symmetric states with maximally mixed one-qubit reductions}

\date{September 19, 2014}

\author{D.~Baguette}
\affiliation{Institut de Physique Nucl\'eaire, Atomique et de Spectroscopie, Universit\'e de Li\`ege, 4000 Li\`ege, Belgium}

\author{T.~Bastin}
\affiliation{Institut de Physique Nucl\'eaire, Atomique et de Spectroscopie, Universit\'e de Li\`ege, 4000 Li\`ege, Belgium}

\author{J.~Martin}
\affiliation{Institut de Physique Nucl\'eaire, Atomique et de Spectroscopie, Universit\'e de Li\`ege, 4000 Li\`ege, Belgium}

\begin{abstract}
We present a comprehensive study of maximally entangled symmetric states of arbitrary numbers of qubits in the sense of the maximal mixedness of the one-qubit reduced density operator. A general criterion is provided to easily identify whether given symmetric states are maximally entangled in that respect or not. We show that these maximally entangled symmetric (MES) states are the only symmetric states for which the expectation value of the associated collective spin of the system vanishes, as well as in corollary the dipole moment of the Husimi function. We establish the link between this kind of maximal entanglement, the anticoherence properties of spin states and the degree of polarization of light fields. We analyze the relationship between the MES states and the classes of states equivalent through stochastic local operations with classical communication (SLOCC). We provide a nonexistence criterion of MES states within SLOCC classes of qubit states and show in particular that the symmetric Dicke state SLOCC classes never contain such MES states, with the only exception of the balanced Dicke state class for even numbers of qubits. The $4$-qubit system is analyzed exhaustively and all MES states of this system are identified and characterized. Finally the entanglement content of MES states is analyzed with respect to the geometric and barycentric measures of entanglement, as well as to the generalized $N$-tangle. We show that the geometric entanglement of MES states is ensured to be larger than or equal to $1/2$, but also that MES states are not in general the symmetric states that maximize the investigated entanglement measures.
\end{abstract}

\maketitle

\section{Introduction}
\label{SecIntroduction}
Entanglement is among the key features of quantum mechanics. It arises when two or more quantum systems interact with each other, even indirectly, and provides nonclassical correlations between them. Entanglement can be used as a resource for various quantum informational tasks such as quantum computation. In the last decades, a lot of effort has been made to quantify the amount of entanglement of various multipartite states, either pure or mixed. This is crucial as a minimal amount of entanglement is needed in pure state quantum computation to outperform classical algorithms~\cite{Linden}. In particular, the search for maximally entangled states (states maximizing certain measures of entanglement) has focused a great deal of attention \cite{Gisin_PLA_246, Hig00,Ost06,Mar10,Aul10}. In the case of $2$ qubits, it is known that Bell states are maximally entangled with respect to any measures of entanglement~\cite{Nielsen}. For higher numbers of qubits, the problem is no longer simple and depends in general on the entanglement measure. In~\cite{Verstraete_PRA_68}, Verstraete \emph{et al.}~refer to maximally entangled states as states with maximally mixed one-qubit reduced density matrices. The same definition was used by Gisin and Bechmann-Pasquinucci~\cite{Gisin_PLA_246}, whereas Scott~\cite{Scott04} uses the term of $1$-uniform states. They are called normal forms in~\cite{Verstraete_PRA_68,Gour-Wallach_NJP_13} and non-generic states in~\cite{Kraus_PRL_104}. These states maximize several measures of entanglement, such as the Meyer-Wallach entanglement measure~\cite{Meyer-Wallach_JMP_43}. They also maximize any entanglement monotone based on linear homogenous positive functions of pure state density matrices within their classes of states equivalent through stochastic local operations with classical communication (SLOCC)~\cite{Verstraete_PRA_68}. Besides, they are conjectured to be maximally entangled with respect to the negative partial transpose measure of entanglement~\cite{Brown_JPA_28}. As appreciated by   Kraus~\cite{Kraus_PRL_104}, they play a specific role in the determination of the local unitary equivalence of multiqubit states. Moreover, they are maximally fragile (in the sense that they are the states which are the most sensitive to noise) and have therefore been proposed as ideal candidates for ultra-sensitive sensors~\cite{Gisin_PLA_246}. All these characteristics together highlight the importance of identifying such maximally entangled states. This problem and its generalization to multiqubit states with maximally mixed $k$-qubit reductions have been approached in~\cite{Arn13,Goy14}. Its complexity grows rapidly with the number of qubits, making analytical results particularly hard to establish. In the case of multiqubit symmetric states, the Hilbert space dimension increases linearly with the number of qubits, which makes the problem easier to tackle. This paper is specifically dedicated to this latter case.
 
The paper is organized as follows. In Sec.~\ref{sec:General-properties-of-MES}, we present a general criterion to quickly identify whether a pure symmetric state of an arbitrary number of qubits is maximally entangled in the sense defined above or not. We then provide two physical interpretations of maximally entangled symmetric (MES) states, one in terms of the collective spin that can be associated with the multiqubit system and a second one in terms of the Husimi function~\cite{Zyczkowski_book} of the state. In Sec.~\ref{sec:On-the-uniqueness}, we study the properties of MES states with respect to local operations assisted with classical communication, the so-called SLOCC operations~\cite{Dur,Benett}. A general non-existence criterion is provided allowing us to know immediately whether SLOCC classes of symmetric states~\cite{Bastin-Krins_PREL_103} can contain MES states or not. An exhaustive identification of all MES states is then performed for 4 qubit systems, after a short reminder of the known $2$- and $3$-qubit cases. In Sec.~\ref{sec:OEC}, we study the entanglement content of MES states with respect to the geometric and barycentric measures of entanglement~\cite{Wei03,Zyczkowski_PRA_85}, as well as to the generalized $N$-tangle~\cite{Wong_PRA_63}. We then draw conclusion in Sec.~\ref{sec:conclusion}. Finally, four appendices about technical results that are used in different sections close this paper.

\section{Maximally entangled symmetric states\label{sec:General-properties-of-MES}}

\subsection{Identification criterion}

The symmetric subspace of an $N$-qubit system gathers all states that are symmetric under any permutation of
the qubits. It is of dimension $N+1$ and is spanned by the orthonormal symmetric Dicke states
\begin{equation}
\label{DNk}
\ket{D_{N}^{(k)}}=\mathcal{N}\sum_{\pi}\ket{\underbrace{0\ldots0}_{N-k}\underbrace{1\ldots1}_{k}}, \quad k = 0, \ldots, N,
\end{equation}
where the sum runs over all permutations of the qubits and $\mathcal{N}$ is a normalization constant. Symmetric Dicke states $\ket{D_{N}^{(k)}}$ are simultaneous eigenstates of $\hat{\mathbf{S}}^{2}$ and $\hat{S}_{z}$ with eigenvalues $N/2(N/2+1)$ and $k-N/2$ where $\hat{\mathbf{S}}$ denotes the collective spin associated to the $N$-qubit system~\cite{footnote1}.

For any $N$-qubit symmetric state $|\psi_S\rangle$, the partial traces over all qubits but $t$ ($0 < t < N$) of $\hat{\rho}_S\equiv|\psi_S\rangle\langle \psi_S|$, $\mathrm{tr}_{\neg t} (\hat{\rho}_S)$, yields identical results for all possible choices of $t$ qubits out of $N$. We can refer in this case to \emph{the} $t$-qubit reduced density operator $\hat{\rho}_t$ of the symmetric state. It reads explicitly (see Appendix A)
\begin{equation}\label{rhotop}
\hat{\rho}_t \equiv \mathrm{tr}_{\neg t}(\hat{\rho}_S)= \sum_{q,\ell=0}^t \bracket{v_t^{(q)}}{v_t^{(\ell)}} \ketbra{D_t^{(q)}}{D_t^{(\ell)}},
\end{equation}
where $|v_t^{(q)}\rangle$ ($q = 0, \ldots, t$) are the $(N-t)$-qubit states
\begin{equation}
\ket{v^{(q)}_t}=\frac{1}{C_{N}^{t}}\sum_{k=0}^{N-t}d_{k+q} \sqrt{C_{N-q-k}^{t-q} C_{k+q}^{k}}\,\ket{D_{N-t}^ {(k)}}.
\end{equation}
Here, $C_i^j$ is the binomial coefficient $\left(\begin{smallmatrix}
i\\j\\
\end{smallmatrix}\right)$ with the usual convention $C_i^j = 0$ for $j < 0$ or $j > i$, and $d_k$ ($k = 0, \ldots, N)$ are the expansion coefficients of the symmetric state $|\psi_S\rangle$ in the Dicke state basis~(\ref{DNk})~:
\begin{equation}
    |\psi_S\rangle = \sum_{k=0}^N d_k |D_N^{(k)}\rangle.
\end{equation}

For instance, the one-qubit reduced density operator $\hat{\rho}_1$ reads in the single-qubit Dicke state basis $\{\ket{D_1^{(0)}}\equiv\ket{0}, \ket{D_1^{(1)}}\equiv\ket{1}\}$
\begin{equation}\label{eq:reduce_density_matrix_1_qubit}
\rho_1=\left(
\begin{array}{cc}
\braket{v^{(0)}_1}{v^{(0)}_1}&\braket{v^{(0)}_1}{v^{(1)}_1}\\[4pt]
\braket{v^{(1)}_1}{v^{(0)}_1}&\braket{v^{(1)}_1}{v^{(1)}_1}\\
\end{array}\right).
\end{equation}

This immediately yields the conditions for a symmetric state to be a MES state, i.e.,
\begin{align}
\braket{v^{(0)}_1}{v^{(0)}_1} & =\sum_{k=0}^{N-1}\frac{N-k}{N}\,|d_{k}|^{2} = \frac{1}{2},\label{NG1}\\
\braket{v^{(1)}_1}{v^{(0)}_1} & =\sum_{k=0}^{N-1}\sqrt{\frac{(N-k)(k+1)}{N^{2}}}\, d_{k}d_{k+1}^{*} = 0,\label{NG2}
\end{align}
or, equivalently, considering the normalization condition,
\begin{align}
& \sum_{k=0}^{N}(N-2k)\,|d_{k}|^{2} = 0,\label{eq:non-genericty-condition-1}\\
& \sum_{k=0}^{N-1}\sqrt{(N-k)(k+1)}\, d_{k}d_{k+1}^{*} = 0.\label{eq:non-genericty-condition-2}
\end{align}

As an example, these conditions show that the symmetric states $(|D_N^{(k)}\rangle + |D_N^{(N-k)}\rangle)/\sqrt{2}$ ($N \geqslant 2$, $k = 0, \ldots, \lfloor N/2 \rfloor - 1$) are MES states~\cite{floorceiling}. For $k=0$, one gets the Bell state ($N=2$) and the $|\mathrm{GHZ}_N\rangle$ states ($N>2$), whose maximal entanglement in that respect is indeed well known~\cite{Gisin_PLA_246}. These states are just a few examples of MES states that can exist for $N$-qubit systems. In Sec.~\ref{sec:On-the-uniqueness} of this paper, an exhaustive analysis of all such states is performed for $N=4$ and some general results are given for arbitrary $N > 4$.

For every $t>1$, the $t$-qubit reduced density operator $\hat{\rho}_t$ is symmetric under any permutation of the $t$ qubits and has only nonzero matrix elements in the $t$-qubit symmetric subspace. We denote hereafter by $\rho_t$ any matrix representation of $\hat{\rho}_t$ in this subspace of dimension $t+1$.

\subsection{Physical Interpretations}

Maximally entangled symmetric (MES) states exhibit interesting properties with respect to the collective spin $\mathbf{S}$ of the system and to the multipole moments of their Husimi functions. These aspects are investigated in the next two subsections.

\subsubsection{In terms of collective spin}

The two following general results hold (see Appendices~\ref{sec:operator_as_pol_spin_col} and ~\ref{sec:reduced_density_matrix_spin_coll})~: any symmetric operator of an $N$-qubit system can be written as a polynomial of degree at most $N$ in the collective spin observables and the matrix elements of the $t$-qubit reduced density matrix $\rho_t$ of any symmetric state $|\psi_S\rangle$ can be written as a polynomial of order $t$ in the expectation values of collective spin observables in the state $|\psi_S\rangle$. 

For instance, the one-qubit reduced density matrix of any symmetric state $|\psi_S\rangle$ [Eq.~(\ref{eq:reduce_density_matrix_1_qubit})] can be reexpressed as
\begin{equation}\label{eq:red_one_spin_coll}
\rho_1=\left(\begin{array}{cc}
\frac{1}{2}- \frac{1}{N}\langle\hat{S}_z \rangle &  \frac{1}{N}\langle \hat{S}_{-} \rangle\\
\frac{1}{N}\langle \hat{S}_{+} \rangle & \frac{1}{2} + \frac{1}{N}\langle \hat{S}_z \rangle
\end{array} \right),
\end{equation}
where the expectation values are meant in the $|\psi_S\rangle$ state.
This merely follows from Eqs.~(\ref{NG1}) and (\ref{NG2}) and the two identities
\begin{eqnarray}
\langle \hat{S}_{z} \rangle & = & \sum_{k=0}^{N}\left(k-N/2\right)\left|d_{k}\right|^{2}\label{eq:mean_Sz},\\
\langle \hat{S}_{+} \rangle & = & \sum_{k=0}^{N-1}\sqrt{(N-k)(k+1)}\,d_{k}d_{k+1}^{*}.
\label{eq:mean_Sp}\end{eqnarray}

Equation~(\ref{eq:red_one_spin_coll}) yields a very interesting physical interpretation of MES states. They are the only states to verify $\langle \hat{S}_{z} \rangle = \langle \hat{S}_{+} \rangle = \langle \hat{S}_{-} \rangle = 0$, or, equivalently,
\begin{equation}\label{eq:spin_col_zero}
\langle \hat{\mathbf{S}}\rangle=0.
\end{equation}
The expectation value of the collective spin vanishes for MES states and only for them. This also implies that these states coincide with spin-$N/2$ order-1 anticoherent states. A spin-$N/2$ state is said to be anticoherent to order $t$ if $\langle (\hat{\mathbf{S}}\boldsymbol{\cdot}\mathbf{n})^k\rangle$ is independent of $\mathbf{n}$ for $k = 1, \ldots, t$, where $\mathbf{n}$ is a unit vector~[\cite{Zimba_EJTP_3}. This definition exactly coincides with Eq.~(\ref{eq:spin_col_zero}) for $t = 1$. Order-1 anticoherence and maximal mixedness of $\hat{\rho}_1$ are thus strictly equivalent concepts. In quantum optics, an analogous concept has been introduced regarding the degree of polarization of light fields~\cite{Luis02}. At the end, order-1 unpolarized light states~\cite{Sanchez-Soto13}, order-1 anticoherent spin states, and $N$-qubit MES states are one and the same concept. Actually, this generalizes to any order~: order-$t$ anticoherence is equivalent to maximal mixedness of $\hat{\rho}_{t}$ in the symmetric subspace~\cite{Giraud}, and consequently to
\begin{equation}
    \rho_{t'} = \frac{1}{t'+1}\idmat_{t'+1}, \quad \forall~t' \leqslant t,
\end{equation}
where $\idmat_{t'+1}$ denotes the $(t'+1)$-dimensional identity matrix.

For instance, the two-qubit reduced density operator $\hat{\rho}_2$ can be expressed in the two-qubit Dicke state basis $\{\ket{D_2^{(0)}}, \ket{D_2^{(1)}}, \ket{D_2^{(2)}}\}$ as
\begin{widetext}\begin{equation}\label{eq:red_two_spin_coll}
\rho_2=\frac{1}{N(N-1)}\left(\begin{array}{ccc}
\langle \hat{S}_{z}^{2}\rangle-\gamma_{N}\langle \hat{S}_{z}\rangle+\beta_{N} & -\sqrt{2}\big(\langle \hat{S}_{z} \hat{S}_{-}\rangle-\alpha_{N}\langle \hat{S}_{-}\rangle\big) & \langle \hat{S}_{-}^{2}\rangle\\
-\sqrt{2}\big(\langle \hat{S}_{+} \hat{S}_{z}\rangle-\alpha_{N}\langle \hat{S}_{+}\rangle\big) & -2\langle \hat{S}_{z}^{2}\rangle+\delta_{N} & \sqrt{2}\big(\langle \hat{S}_{z} \hat{S}_{-}\rangle+{\alpha_{N}}\langle \hat{S}_{-}\rangle\big)\\
\langle \hat{S}_{+}^{2}\rangle & \sqrt{2}\big(\langle \hat{S}_{+} \hat{S}_{z}\rangle+{\alpha_{N}}\langle \hat{S}_{+}\rangle\big) & \langle \hat{S}_{z}^{2}\rangle+\gamma_{N}\langle \hat{S}_{z}\rangle+\beta_{N}
\end{array}\right),
\end{equation}\end{widetext}
where $\alpha_{N}=\sN-1$, $\beta_{N}=\sN(\sN-1)$, $\gamma_{N}=2\sN-1$, and $\delta_{N}=2\sN^2$, with $s_N = N/2$. Furthermore, order-2 anticoherence is equivalent to the conditions~\cite{Bannai_JPA_44}
\begin{equation}\label{eq:anticoherent2_spin}
\begin{aligned}
&\langle\hat{S}_x\rangle=\langle\hat{S}_y\rangle = \langle\hat{S}_z\rangle = 0, \\
&\langle\hat{S}_x\hat{S}_y\rangle=\langle\hat{S}_y\hat{S}_z\rangle=\langle\hat{S}_z\hat{S}_x\rangle=0,\\
&\langle\hat{S}_x^2\rangle=\langle\hat{S}_y^2\rangle=\langle\hat{S}_z^2\rangle,
\end{aligned}
\end{equation}
which leads to $\rho_2=\idmat_3/3$ and $\rho_1=\idmat_2/2$.

The $t$-qubit reduced density matrix $\rho_t$ of any pure $N$-qubit symmetric state has at most $\min(t+1,N-t+1)$ nonvanishing eigenvalues. Indeed, this follows from Schmidt decomposition which implies that $\rho_t$ and $\rho_{N-t}$ have the same spectra, aside from zeros~\cite{Gisin_PLA_246}. In order to be anticoherent to order $t$, the $t$-qubit reduced density matrix must be full rank which is only possible if $t \leqslant N/2$. As a consequence, any pure symmetric state of $N$ qubits can be anticoherent at most to order $\lfloor N/2 \rfloor$.

\subsubsection{In terms of multipole moments of the Husimi function}

We now turn to another interpretation of MES states relying on the Husimi function. From the continuous set of separable states $\ket{\Phi(\theta,\varphi)}=\ket{\phi(\theta,\varphi)}^{\otimes N}$ with $\ket{\phi(\theta,\varphi)}=\cos(\theta/2)\ket{0}+e^{i\varphi}\sin(\theta/2)\ket{1}$, one constructs the Husimi function of a symmetric state $\ket{\psi_S}$ as~\cite{Zyczkowski_book}
\begin{equation}
\label{HusDistr}
\mathcal{H}(\theta,\varphi)=|\bracket{\Phi(\theta,\varphi)}{\psi_S}|^2.
\end{equation}
The Husimi function is a quasiprobability distribution on the sphere verifying the normalization condition
\begin{equation}
\frac{N+1}{4 \pi}\int \mathcal{H}(\theta,\varphi)\,d\Omega = 1.
\end{equation}

The following interpretation of MES states can then be given : a symmetric state is a MES state iff the dipole moment of its Husimi function vanishes, i.e., iff
\begin{equation}
    \mathbf{d} \equiv \int \mathbf{r}\,\mathcal{H}(\theta,\varphi)\,d\Omega = 0,
\end{equation}
where $\mathbf{r}=(\sin\theta\cos\varphi, \sin\theta\sin\varphi, \cos\theta)$.
The proof relies on the expression of the collective spin operators in the overcomplete separable state basis,
\begin{equation}
\hat{S}_i=\mathcal{K}_N\int r_i \,\ket{\Phi(\theta, \varphi)}\bra{\Phi(\theta, \varphi)}\,d\Omega,
\end{equation}
with $i=x,y,z$ and $\mathcal{K}_N = (N+1)(N+2)/8 \pi$~\cite{Lieb73,Kut73}. From this representation, the expectation value of the spin operator readily follows,
\begin{equation}\label{dipmom}
\langle \hat{\mathbf{S}} \rangle = \mathcal{K}_N \int \mathbf{r}\,\mathcal{H}(\theta,\varphi)\,d\Omega = \mathcal{K}_N \mathbf{d}.
\end{equation}
But as shown previously, a state is a MES state (or anticoherent state to order 1) iff $\langle \hat{\mathbf{S}}\rangle=0$, that is iff $\mathbf{d} = 0$. This interpretation can be pursued to higher orders of anticoherence. An anticoherent state to order 2 will be characterized by vanishing dipole and quadrupole moments of its Husimi function. Indeed, a little algebra shows that the expectation values $\langle  \hat{S}_j \hat{S}_k \rangle$ translate into
\begin{equation}\label{eq:spinmean_quadripolar1}
\langle  \hat{S}_j \hat{S}_k \rangle = \mathcal{N}_N\int \left[r_j r_k + \frac{i \sum_l \epsilon_{j k l}\,r_l + \delta_{jk}}{N+3}\right] \mathcal{H}(\theta,\varphi)\,d\Omega
\end{equation}
with $\mathcal{N}_N=(N+1)\left(N+2\right)\left(N+3\right)/16\pi$ and where $\epsilon_{j k l}$ is the Levi-Civita symbol. It is now easy to see from Eqs. (\ref{dipmom}) and (\ref{eq:spinmean_quadripolar1}) that the conditions (\ref{eq:anticoherent2_spin}) of anticoherence to order 2 are satisfied iff the dipole and the quadrupole moments of the Husimi function vanish, i.e., iff $\mathbf{d}=0$ and 
\begin{equation}
{Q}_{j k}=\int (3 r_j r_k -  r^2 \delta_{j k})\,\mathcal{H}(\theta,\varphi)\,d\Omega = 0,\;\;\;\forall\,j,k.
\end{equation}

As these developments suggest, a much more general result holds : A state $\ket{\psiS}$ is anticoherent to order $t$ iff all multipole moments up to order $2^t$ of its Husimi function vanish~\cite{Giraud}.

\section{Maximally Entangled Symmetric States and SLOCC
classes\label{sec:On-the-uniqueness}}

General $N$-qubit systems are known to exhibit several types of entanglement with respect to the stochastic interconvertibility of the states through local operations with classical communication, the so-called SLOCC operations~\cite{Dur}. This entanglement richness is reflected in the $N$-qubit Hilbert space by the SLOCC classes gathering together all states interconvertible to each other through these operations. MES states are not found within each SLOCC class. However, should there be, they are unique up to local unitaries~\cite{Verstraete_PRA_68,Gour-Wallach_NJP_13}. In this section, we provide a general nonexistence criterion of MES states in SLOCC classes and we explicitly identify all of them in the $4$-qubit case.

The SLOCC classes in the symmetric subspace have been described in Ref.~\cite{Bastin-Krins_PREL_103}. They follow the Majorana representation that writes the symmetric states as
\begin{equation}
\label{mr}
\ket{\psiS}=\mathcal{N}\sum_{\pi}\ket{\phi_{\pi(1)}\ldots\phi_{\pi(N)}},
\end{equation}
where $|\phi_1\rangle, \ldots, |\phi_N\rangle$ are single-qubit states, the sum runs over all permutations $\pi$ of the qubits, and $\mathcal{N}$ is a normalization constant. In this form, symmetric states can be geometrically represented by $N$ points on the Bloch sphere, the so-called Majorana points, associated with the individual single-qubit states $|\phi_i\rangle$. Some of these points can be superimposed on each other (in the case of equality of the corresponding individual states), yielding a single distinct point on the Bloch sphere with a multiplicity larger than 1. The total number of distinct points defines the diversity degree $d$ of the symmetric states $|\psi_S\rangle$ and the list $\ell$ of multiplicities of each distinct point, sorted by decreasing order, defines the degeneracy configuration $\mathcal{D}_{\ell}$ of the states. All states with identical such parameters are gathered into state families denominated accordingly. In the symmetric subspace, SLOCC classes contain only states with identical degeneracy configurations and the number of SLOCC classes of states of a given degeneracy configuration $\mathcal{D}_{\ell}$ is either 1 (if $d \leqslant 3$) or infinite (if $d>3$)~\cite{Bastin-Krins_PREL_103}. In the first case, the SLOCC classes can be unambiguously denominated by the degeneracy configuration of the states they gather. In particular, the SLOCC classes $\mathcal{D}_N$ gather all $N$-qubit separable states and the SLOCC classes $\mathcal{D}_{N-k,k}$ ($k = 1, \ldots, \lfloor N/2 \rfloor)$ gather for each $k$ all states that are SLOCC equivalent to the Dicke states $|D_N^{(k)}\rangle$~\cite{Bastin-Krins_PREL_103}.

\textbf{Non-existence criterion~:} SLOCC classes of $\mathcal{D}_{\ell}$-type symmetric states with $\ell$ containing a multiplicity $m \ge N/2$ do not contain any MES states, except for the $\mathcal{D}_{N/2,N/2}$ SLOCC class when $N$ is even, in which case $\ket{D_{N}^{(N/2)}}$ is such a state.

Indeed, any symmetric state with $m$ identical single-qubit states $|\phi_i\rangle$ in the Majorana representation (\ref{mr}) can be mapped through local unitaries to a symmetric state with $m$ $|\phi_i\rangle$ equal to $|0\rangle$ (the local unitaries $U^{\otimes N}$ with $U |\phi_i\rangle = |0\rangle$ are convenient for this purpose). Such a transformed state is a linear superposition of multiqubit states with at most $N-m$ excitations $|1\rangle$ and has thus no components $d_k$ on any Dicke states $|D_N^{(k)}\rangle$ with $k > N-m$. If $m \ge N/2$, $d_k = 0$, for all $k > N/2$, and the left-hand side of Eq.~(\ref{eq:non-genericty-condition-1}) is strictly positive, unless $d_{N/2}$ is the only coefficient to be nonzero, in which case both Eqs.~(\ref{eq:non-genericty-condition-1}) and ~(\ref{eq:non-genericty-condition-2}) are satisfied. In the first case, Eq.~(\ref{eq:non-genericty-condition-1}) can never be satisfied and the symmetric state can never be maximally entangled. This ends the proof. For all SLOCC classes not addressed by our criterion, a general statement about the existence of MES states remains an open problem and each case must be considered individually.

As a consequence of our criterion, and as can also be inferred from the recent work of Walter~\emph{et~al.}~\cite{Wal13} on entanglement polytopes, the balanced Dicke state SLOCC class $\mathcal{D}_{N/2,N/2}$ ($N$ even) is the only Dicke state class that contains a MES state (up to local unitary), all others, i.e., the classes $\mathcal{D}_{N-k,k}$ ($k = 1, \ldots, \lceil N/2 \rceil -1 $), do not contain any. With the exception of the balanced Dicke state case, this statement generalizes to arbitrary $N$ and $k$ in the symmetric subspace the result of Verstraete~\textit{et al.}~\cite{Verstraete_PRA_68}, according to which the 3-qubit $W$ SLOCC class (i.e., the Dicke state $|D_3^{(1)}\rangle$ class) does not contain any maximally entangled states. Incidentally, our criterion also states that the $\mathcal{D}_N$ SLOCC classes do not contain any MES states, but this case is obvious since these classes correspond for each $N$ to the separable state classes.

Specifically for $N = 2$, there are only the 2 SLOCC classes $\mathcal{D}_2$ (separable states) and $\mathcal{D}_{1,1}$ (entangled states)~\cite{Bastin-Krins_PREL_103}. According to our criterion, the $\mathcal{D}_2$ class does not contain any MES states, while in $\mathcal{D}_{1,1}$ the Bell state $|D_1^{(1)}\rangle \equiv (|10\rangle + |01\rangle)/\sqrt{2}$ is such a one. Both cases are obvious. For $N = 3$, there are only the 3 SLOCC classes $\mathcal{D}_3$ (separable states), $\mathcal{D}_{2,1}$ (W class), and $\mathcal{D}_{1,1,1}$ (GHZ class)~\cite{Bastin-Krins_PREL_103}. According to our criterion, $\mathcal{D}_3$ and $\mathcal{D}_{2,1}$ cannot contain any MES states. In the first case, this is obvious, in the second case, this was shown by Verstraete~\textit{et al.}~\cite{Verstraete_PRA_68}. The last class $\mathcal{D}_{1,1,1}$ contains the $3$-qubit GHZ state, which is known to be maximally entangled independently of the number of qubits~\cite{Gisin_PLA_246}. The Majorana representation of the GHZ state consists of 3 points angularly equally spaced on the equatorial plane of the Bloch sphere.

\begin{figure}
\begin{centering}
\includegraphics[width=0.26\textwidth,clip=true]{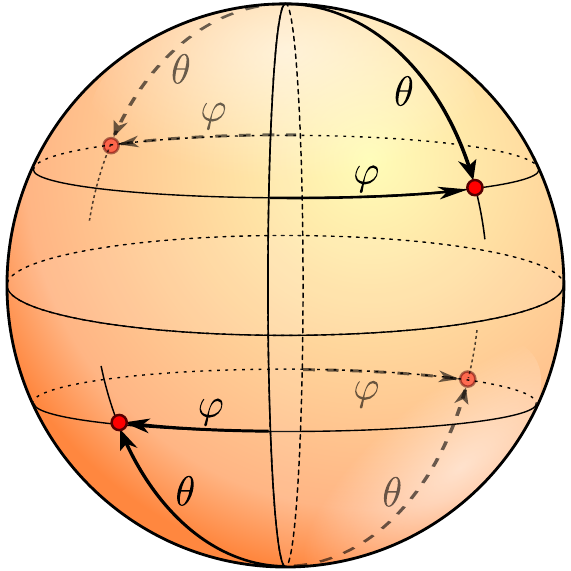}
\par\end{centering}
\caption{(Color online) Majorana representation of the MES states $|\psi_{\mu}\rangle$ [Eq.~(\ref{nongen4})] belonging to the 4-qubit SLOCC classes $\mathcal{C}_{\mu}^{1,1,1,1}$ [$\mu \in \mathit{S}$, $(\theta,\varphi) \in \mathit{S}'$].\label{fig:MES-of-4-qub}}
\end{figure}

For $N=4$, we have the 4 SLOCC classes $\mathcal{D}_4$ (separable states), $\mathcal{D}_{3,1}$ (class of the W state $|D_4^{(1)}\rangle$), $\mathcal{D}_{2,2}$ (class of the balanced Dicke state $|D_4^{(2)}\rangle$) and $\mathcal{D}_{2,1,1}$, as well as the infinite number of SLOCC classes of the $\mathcal{D}_{1,1,1,1}$ state family~\cite{Bastin-Krins_PREL_103}. According to our criterion, the classes $\mathcal{D}_4$, $\mathcal{D}_{3,1}$ and $\mathcal{D}_{2,1,1}$ do not contain any MES states, contrary to the $\mathcal{D}_{2,2}$ class where the balanced Dicke state $|D_4^{(2)}\rangle$ is a representative. In the $\mathcal{D}_{1,1,1,1}$ state family, all symmetric states are SLOCC equivalent to one of the states (see Appendix~\ref{app:slocc4qubit})
\begin{equation}
\label{nongen4}
\ket{\psi_\mu}=\frac{1}{\sqrt{2 + |\mu|^2}}\left(\ket{D_{4}^{(0)}}+\mu\ket{D_{4}^{(2)}}+\ket{D_{4}^{(4)}}\right)
\end{equation}
with $\mu$ a $c$-number in the bounded domain
\begin{align}
\label{Ddef}
    \mathit{S} = & \{ \mu \in \mathbb{C} : \Re(\mu) \geqslant 0, \Im(\mu) \geqslant 0, \nonumber \\
    & |\mu - \sqrt{2/3}| \leqslant 2 \sqrt{2/3}, \mu < \sqrt{2/3} \textrm{ if } \Im(\mu) = 0\}.
\end{align}
In particular, $|\psi_{\mu = 0}\rangle = |\mathrm{GHZ}_4\rangle$. In Eq.~(\ref{nongen4}) states with different $\mu \in \mathit{S} $ are SLOCC inequivalent (see Appendix~\ref{app:slocc4qubit}). All SLOCC classes of the $\mathcal{D}_{1,1,1,1}$ state family can thus be unambiguously identified by this $c$-number and denoted accordingly by $\mathcal{C}_{\mu}^{1,1,1,1}$ with $|\psi_{\mu}\rangle$ as a representative. All these classes admit MES states since so are the representatives $|\psi_{\mu}\rangle$ which verify Eqs.~(\ref{eq:non-genericty-condition-1}) and (\ref{eq:non-genericty-condition-2}). The Majorana representation of these states consists of 4 distinct points on the Bloch sphere with polar and azimuthal coordinates $(\theta,\varphi)$, $(\theta,\pi+\varphi)$, $(\pi-\theta,-\varphi)$, $(\pi-\theta,\pi-\varphi)$, such that $\mu=-(z^2+1/z^2)/\sqrt{6}$ with $z=\cot(\theta/2)\,e^{-i\varphi}$ (see Appendix~\ref{app:slocc4qubit}). This Majorana representation is shown in Fig.~\ref{fig:MES-of-4-qub}. It exhibits interestingly the dihedral $D_2$ point group symmetry. The domain of the angular coordinates $(\theta,\varphi)$ in bijection with the domain $\mathit{S}$ of the $c$-numbers $\mu$ is given by
\begin{equation}
    \mathit{S}' = ]\pi/4,\pi/2] \times [\varphi_{\mathrm{min}}(\theta),\pi/2[
\end{equation}
with $\varphi_{\mathrm{min}}(\theta) = \max(\pi/4,\arcsin(\cot \theta))$. On the Bloch sphere, this restricted domain is delimited by the meridian planes $\varphi = \pi/4$ and $\varphi = \pi/2$, the equatorial plane $\theta = \pi/2$, and the oblique plane passing through the points $(\pi/2, 0)$, $(\pi/4, \pi/2)$ and $(\pi/2, \pi)$.

\begin{figure}
\begin{centering}
\includegraphics[width=0.38\textwidth]{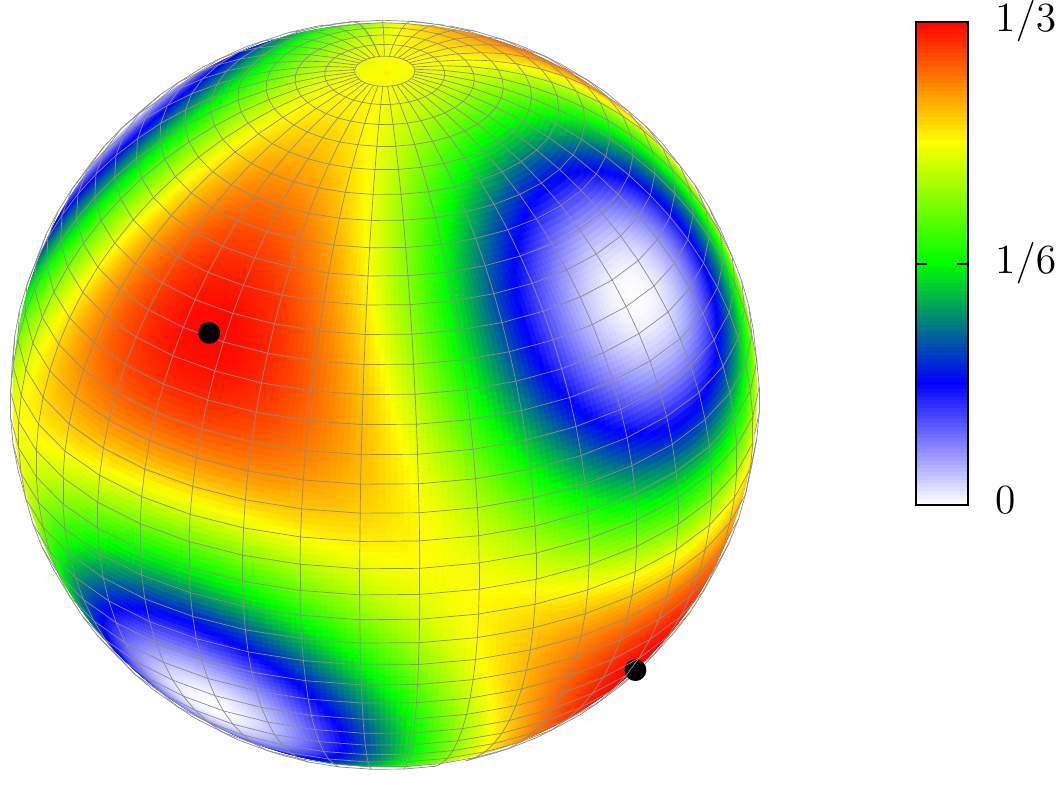}
\par\end{centering}
\caption{(Color online) Density plot of the Husimi function (\ref{HusDistr}) associated with the tetrahedron state $|T_4\rangle \equiv |\psi_{\mu = i \sqrt{2}}\rangle$ (black dots are the Majorana points of the state, two being only visible in the picture). \label{fig:Husimi}}
\end{figure}

The two-qubit reduced density matrices of the states $|\psi_{\mu}\rangle$ read explicitly
\begin{equation}\label{rho2mu}
\rho_2=\frac{1}{2+|\mu|^{2}}\left(\begin{array}{ccc}
\displaystyle
1+\frac{|\mu|^{2}}{6} & 0 & \displaystyle\sqrt{\frac{2}{3}}\,\Re(\mu)\\
0 & \displaystyle\frac{2}{3}|\mu|^{2} & 0\\
\displaystyle\sqrt{\frac{2}{3}}\,\Re(\mu)& 0 & \displaystyle 1+\frac{|\mu|^{2}}{6}
\end{array}\right).
\end{equation}
Only the state $|\psi_{\mu}\rangle$ with $\mu = i\sqrt{2}$ is anticoherent to order 2. This is even the only $4$-qubit state to be so since the two-qubit reduced density matrix of the balanced Dicke state $|D_4^{(2)}\rangle$ is equal to $\mathrm{diag}(1/6,2/3,1/6) \neq \idmat_3/3$. The state $|T_4\rangle \equiv |\psi_{\mu = i\sqrt{2}}\rangle$ is the $4$-qubit tetrahedron state (the 4 Majorana points draw a regular tetrahedron)~\cite{Mar10}. Figure~\ref{fig:Husimi} shows a density plot of the Husimi function associated with this state, which is characterized by vanishing dipole and quadrupole moments.

\section{Entanglement content of maximally entangled symmetric states}
\label{sec:OEC}

Maximally entangled symmetric (MES) states as defined in this work maximize many measures of entanglement (see Introduction) but not all. In particular, the geometric and barycentric measures of entanglement~\cite{Wei03, Zyczkowski_PRA_85}, or the generalized $N$-tangle~\cite{Wong_PRA_63}, to cite a few, are not maximized for all MES states. In this section, we address in more details this question of the entanglement content of MES states with respect to these entanglement measures.

\subsection{Geometric measure of entanglement}

The geometric measure of entanglement (GME) $E_G$ of a state $\ket{\psi}$ is defined as~\cite{Wei03}
\begin{equation}\label{geoent}
E_G(\psi) \equiv 1-\max_{\ket{\Phi} = \ket{\phi_1,\phi_2,\phi_3,\ldots}} |\bracket{\Phi}{\psi}|^2.
\end{equation}
If $|\psi\rangle$ is a symmetric state, the optimization can be done on the limited set of symmetric separable states $\ket{\Phi}=|\phi, \ldots, \phi\rangle$~\cite{Hub09}. The geometric measure of entanglement $E_G$ of any $N$-qubit MES state is ensured to be larger than or equal to $1/2$, the equality only holding for Bell states ($N=2$), GHZ states ($N>2$), and their local unitarily (LU) equivalents~:
\begin{equation}
\label{EG}
E_G(\psi_{\mathrm{MES}}) \geqslant \frac{1}{2}, \quad \forall \textrm{ MES state } \ket{\psi_{\mathrm{MES}}}.
\end{equation}
Indeed, any symmetric separable states can be obtained from the action of a local unitary $\hat{U}^{\otimes N}$ on the separable state $\ket{0,\ldots,0}\equiv \ket{D_N^{(0)}}$. We thus have
\begin{equation}\label{maxoverlap}
\max_{\ket{\Phi} = \ket{\phi, \ldots, \phi}} |\bracket{\Phi}{\psi}|^2
= \max_{\hat{U} \in U(2)} |\bracket{D_N^{(0)}}{\hat{U}^{\otimes N}|\psi}|^2.
\end{equation}
When $\ket{\psi}$ is a MES state, so is $\hat{U}^{\otimes N}\ket{\psi}$, and it follows from Eq.~(\ref{NG1}) that $|\tilde{d}_0|^2\leqslant 1/2$ where $\tilde{d}_0=\bracket{D_N^{(0)}}{\hat{U}^{\otimes N}|\psi}$. This shows that the geometric entanglement of $\hat{U}^{\otimes N}\ket{\psi}$ and hence $\ket{\psi}$ is necessarily larger than or equal to $1/2$. When $|d_0|^2 = 1/2$,  Eq.~(\ref{NG1}) shows that all $d_k$ must vanish for $0<k\leqslant N-1$ and normalization imposes $|d_N|^2 = 1/2$, which eventually leads to the Bell state ($N=2$) or GHZ states ($N>2$), up to local unitaries.

Equation~(\ref{EG}) can be generalized to higher order of anticoherence. The geometric measure of entanglement of any anticoherent state to order $t$, $\ket{\psi_{\mathrm{A}}^{(t)}}$, is larger than or equal to $t/(t+1)$~:
\begin{equation}
    E_G(\psi_{\mathrm{A}}^{(t)}) \geqslant \frac{t}{t+1}, \quad \forall~\ket{\psi_{\mathrm{A}}^{(t)}}.
\end{equation}
Indeed, from the condition $\langle v_t^{(0)}|v_t^{(0)} \rangle = 1/(t+1)$, we find that the modulus of $d_0$ must satisfy the equation
\begin{equation}
\frac{1}{t+1}=|d_0|^2+\frac{1}{C_N^t}\sum_{k=1}^{N-t} |d_k|^2 C_{N-k}^t,
\end{equation}
which immediately leads to $|d_0|^2\leqslant 1/(t+1)$ and $E_G\geqslant t/(t+1)$.

For $N=2$, the only MES state (up to LU) is the Bell state with a GME of $1/2$. This is the maximal value that can be obtained for 2-qubit states~\cite{Wei03}. For $N=3$, the only MES state is the GHZ state and its GME is also $1/2$~\cite{Wei03}. However, in this case it doesn't maximize the GME since the maximal value for $3$-qubit states is obtained for the Dicke state $|D_3^{(1)}\rangle$ with a GME of 5/9~\cite{Chen}. The GHZ state doesn't maximize either the GME within its SLOCC class since this maximal value is also 5/9 as can be inferred from \cite{Acin} (the Dicke state $|D_3^{(1)}\rangle$ can be approached as closely as desired by GHZ-class states).

\begin{figure}
\begin{centering}
\includegraphics[width=0.47\textwidth]{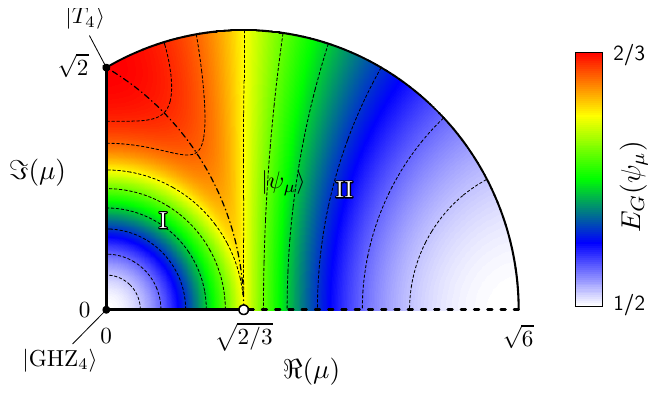}
\par\end{centering}
\caption{(Color online) Density plot of the geometric measure of entanglement $E_G$ of the 4-qubit MES states $|\psi_{\mu}\rangle$ [Eq.~(\ref{nongen4})] as a function of the real and imaginary parts of $\mu \in \mathit{S}$ [Eq.~(\ref{Ddef})], the only region where distinct $\mu$ define SLOCC inequivalent $\mathcal{D}_{1,1,1,1}$ states (see text). The geometric entanglement is constant along the dashed curves. These curves cross the region boundaries at right angles. Particular values of $\mu$ are highlighted~: $|\psi_{\mu = 0}\rangle = |\mathrm{GHZ}_4\rangle$, $|\psi_{\mu = i\sqrt{2}}\rangle \equiv |T_4\rangle$ (tetrahedron state). The dash-dotted arc of circle separates two regions I and II where the density plots of $E_G$ are just the image of each other through the conjugated Moebius transformation $\mu \rightarrow [2(\sqrt{6} - \mu)/(\sqrt{6} \mu + 2)]^{\ast}$ (see text).\label{fig:EG}}
\end{figure}

For $N=4$, the only MES states are the balanced Dicke state $|D_4^{(2)}\rangle$ and the states $|\psi_{\mu}\rangle$ given by Eq.~(\ref{nongen4}). The GME of the balanced Dicke state is $5/8$~\cite{Wei03} while it can be expressed for the states $|\psi_{\mu}\rangle$ as
\begin{equation}\label{squared_overlap}
E_G(\psi_{\mu}) = 1 - \max_{\theta \in [0,\pi], \varphi \in [0,2\pi]} \frac{\big| \alpha^4+\mu\sqrt{6}\,\alpha^2\beta^2+\beta^4 \big|^2}{2+|\mu|^2}
\end{equation}
with $\alpha = \cos(\theta/2)$ and $\beta=\sin(\theta/2)e^{i\varphi}$. The GME of the states $|\psi_{\mu}\rangle$ is represented in Fig.~\ref{fig:EG} for all $\mu \in \mathit{S}$ [Eq.~(\ref{Ddef})], the only region where distinct $\mu$ define SLOCC inequivalent $\mathcal{D}_{1,1,1,1}$ states (see Sec.~\ref{sec:On-the-uniqueness}). For $\mu$ such as $|\mu| \leqslant \sqrt{2/3}$, $E_G(\psi_{\mu}) = (1+|\mu|^2)/(2+|\mu|^2)$ and $\ket{\Phi}=\ket{0}^{\otimes N}$ is a separable state maximizing the squared overlap in Eq.~(\ref{geoent}). For $\mu$ such that $\Re(\mu) = 0$ and $|\mu| > \sqrt{2/3}$, $E_G(\psi_{\mu}) = 1 - (2 + 3 |\mu|^2)^2/[24 |\mu|^2 (2 + |\mu|^2)]$ and $\ket{\Phi_\mu}=(\alpha_\mu\ket{0}+\beta_\mu\ket{1})^{\otimes N}$ with $\alpha_\mu=[1/2+1/(\sqrt{6}|\mu|)]^{1/2}$ and $\beta_\mu=e^{i\pi/4}[1/2-1/(\sqrt{6}|\mu|)]^{1/2}$ is a separable state maximizing the squared overlap in Eq.~(\ref{geoent}). For any $\mu$ in region I of Fig.~\ref{fig:EG}, $|\psi_{\mu}\rangle$ and $|\psi_{\mu'}\rangle$ with $\mu' = 2 (\sqrt{6} - \mu)/(\sqrt{6} \mu + 2)$ are LU-equivalent and the transformation $\mu'(\mu)^{\ast}$ maps region I to region II and vice versa (see Appendix \ref{app:slocc4qubit}). Since the geometric entanglement is invariant through LU and complex conjugation, we have $E_G(\psi_{\mu}) = E_G(\psi_{\mu'^{\ast}})$ and the density plot of region II is just the image of region I's through the anticonformal transformation $\mu'(\mu)^{\ast}$. The boundary between regions I and II is defined by the arc of circle of radius $2\sqrt{2/3}$ centered in $-\sqrt{2/3}$. Along this arc, the GME reads
$E_G(\psi_{\mu}) = 1 - (2 + 3 |\mu''|^2)^2/[24 |\mu''|^2 (2 + |\mu''|^2)]$ with $\mu'' = 2 (\sqrt{6} + \mu)/(\sqrt{6} \mu - 2)$. The maximal GME of the 4-qubit MES states $|\psi_{\mu}\rangle$ is reached for $\mu=i\sqrt{2}$ (tetrahedron state) with $E_G=2/3$. This is actually the maximum GME that can be achieved for 4-qubit symmetric states~\cite{Mar10,Aul10}. This is also incidentally the only 4-qubit state that is anticoherent to order 2 (see Sec.~\ref{sec:On-the-uniqueness}).

\subsection{Barycentric measure of entanglement}

\begin{figure}
\begin{centering}
\includegraphics[width=0.47\textwidth]{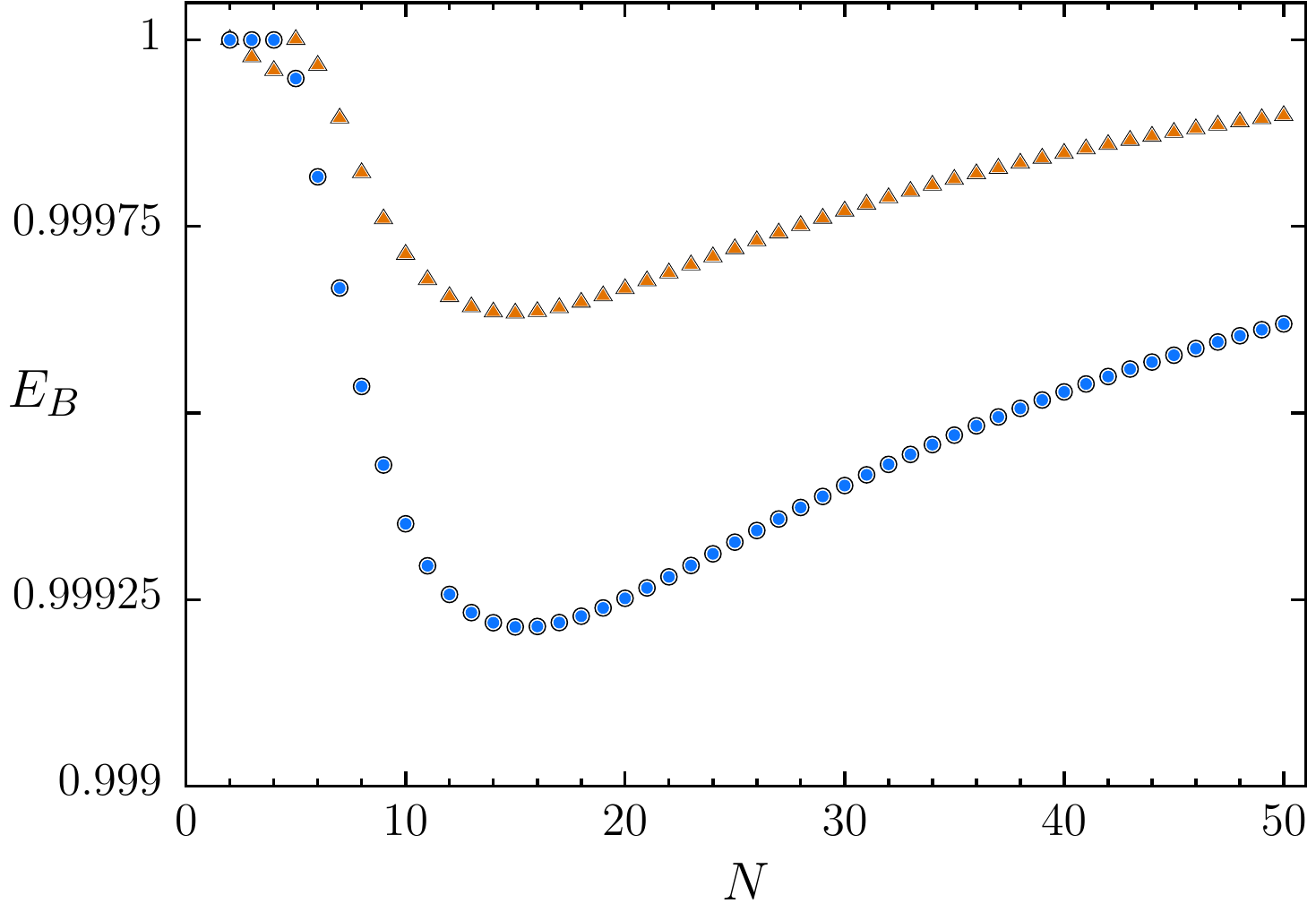}
\par\end{centering}
\caption{(Color online) Barycentric measure of entanglement $E_B$ of the MES states $|P_N\rangle$ (blue dots) and the SLOCC equivalent non-MES states $|P_{N,\alpha}\rangle$ ($\alpha = -\sqrt[8]{27/25}$) (orange triangles) as a function of the number of qubits $N$.\label{fig:Bgraph}}
\end{figure}

The barycentric measure of entanglement (BME) $E_B$ of a symmetric state $|\psi_S\rangle$ is defined as~\cite{Zyczkowski_PRA_85}
\begin{equation}
E_B(\psi_S) \equiv 1-d_B^2(\psi_S),
\end{equation}
where $d_B(\psi_S)$ is the Euclidian distance from the Bloch sphere center to the barycenter of the Majorana points of $\ket{\psi_S}$. All MES states up to 4 qubits (see Sec.~\ref{sec:On-the-uniqueness}) have Majorana points with a barycenter that coincides with the Bloch sphere center. These states are therefore maximally entangled with respect to the BME ($E_B=1$). This is no longer true when considering states of more than $4$ qubits. Still MES states with maximal BME of 1 can be found for any numbers of qubits (such as the GHZ states $\ket{\mathrm{GHZ}_N}$ whose Majorana points draw a regular $N$-sided polygon in the equatorial plane~\cite{Bastin-Krins_PREL_103}), but MES states with smaller BME can also be identified for any $N > 4$. Some were already pointed out for $N=5$ and $N=7$ in \cite{Bannai_JPA_44} as anticoherent states whose Majorana points do not define a spherical $1$-design. We identify here a series of such MES states for any numbers of qubits larger than 4. The states
\begin{equation}\label{eq:BCN_state}
\ket{P_N}=\frac{1}{\sqrt{2N-2}}\big(\sqrt{N-2}\,\ket{D_{N}^{(0)}}+\sqrt{N}\,\ket{D_{N}^{(N-1)}}\big)
\end{equation}
are MES for any $N \geqslant 2$ [they indeed verify Eqs.~(\ref{eq:non-genericty-condition-1}] and (\ref{eq:non-genericty-condition-2})). Their Majorana representations are formed with one point at the north pole of the Bloch sphere and $N-1$ points at the vertices of a regular polygon contained in a plane parallel to the equatorial plane but slightly displaced towards the south pole, with a polar angle $\theta=2\sqrt[N-1]{\cot^{-1}[(N-2)/N^2]}$. The BME of the states $|P_N\rangle$ reads accordingly
\begin{equation}\label{BCN}
E_B(P_N)=1-\left[\frac{2(N-1)}{N(1+\sqrt[N-1]{(N-2)/N^2})}-1\right]^2.
\end{equation}
It is illustrated in Fig.~\ref{fig:Bgraph} as a function of $N$. The curve is slightly below 1 for any $N > 4$ and displays a minimum for $N=15$. For very large $N$, $E_B(P_N)$ tends again to 1 while staying smaller. For $N \leqslant 4$, the states $\ket{P_N}$ identify to the Bell state ($N=2$), or to LU-equivalent states to the $|\mathrm{GHZ}_3\rangle$ state ($N=3$) or to the tetrahedron state $|T_4\rangle \equiv |\psi_{\mu = i \sqrt{2}}\rangle$ ($N = 4$). Incidentally, the geometric entanglement of the states $|P_N\rangle$ is equal to $N/(2N-2)$ for $N \geqslant 4$ and $1/2$ otherwise.

Conversely, states that maximize the barycentric measure of entanglement ($E_B=1$) are not necessarily MES states. For instance, the states ($N > 2$)
\begin{eqnarray*}
\ket{\chi_N}&=&\mathcal{N}\bigg(\sqrt{\frac{C_N^2}{3}}\ket{D_{N}^{(0)}}-\sqrt{3}\ket{D_{N}^{(2)}}+\frac{(-1)^N}{\sqrt{N-2}}\ket{D_{N}^{(N-3)}}\\
& & + (-1)^{N+1}\sqrt{\frac{3 N-3}{2}}\ket{D_{N}^{(N-1)}}\bigg)
\end{eqnarray*}
have maximal BME but are not MES for $N > 4$. Indeed, their Majorana representations correspond to $3$ points at the vertices of an equilateral triangle in the meridian plane $\varphi = 0$ and $N-3$ points at the vertices of a regular polygon contained in the equatorial plane, such that $d_B=0$ and $E_B=1$. Furthermore, the first element of their one-qubit reduced density matrix reads
\begin{equation}
\braket{v^{(0)}_1}{v^{(0)}_1}=\frac{N^4 - 3 N^3 + 29 N^2 - 99 N + 108}{N (N + 1) (N^2 + 5 N - 12)}.
\end{equation}
This element is equal to $1/2$ only for $N=3$ and $N=4$ and according to Eq.~(\ref{NG1}) the state cannot be a MES state for $N > 4$.

MES states do not generally maximize the BME within their SLOCC classes. For instance, the $|P_N\rangle$ SLOCC-equivalent states ($\alpha \in \mathbb{C}_0$)
\begin{align}\label{eq:SLOCC}
\ket{P_{N,\alpha}} & = \mathcal{N}[\mathrm{diag}(\alpha, 1)]^{\otimes N} |P_N\rangle \\
& = \mathcal{N}'\big(\alpha^{N-1}\sqrt{N-2}\,\ket{D_{N}^{(0)}}+\sqrt{N}\,\ket{D_{N}^{(N-1)}}\big) \nonumber
\end{align}
are non-MES states as long as $|\alpha| \neq 1$ [since $\mathrm{diag}(\alpha, 1)$ is non-unitary in this case] and though they exhibit a larger BME than the $|P_N\rangle$ states for $N > 4$ and several values of $\alpha$, in particular for $\alpha =- \sqrt[8]{27/25}$ as is illustrated in Fig.~\ref{fig:Bgraph}.

\subsection{Generalized $N$-tangle}
The generalized $N$-tangle $\tau_N$ of a state $|\psi\rangle$ was introduced in \cite{Wong_PRA_63} as a measure of multipartite entanglement. For all even $N$, it is equal to the square of the concurrence~:
\begin{equation}
\tau_N(\psi)=|\bra{\psi}{\sigma_y^{\otimes N}\ket{\psi^*}}|^4
\end{equation}
with $\sigma_y$ the second Pauli matrix. For $N=3$, it corresponds to the usual $3$-tangle~\cite{Dur}. The MES states of $2$ and $3$ qubits have maximal $N$-tangle $\tau_N=1$. From $N = 4$, the $N$-tangle of MES states can span values between 0 and 1. We have in particular for the $4$-qubit states $|\psi_\mu\rangle$ of Eq.~(\ref{nongen4})
\begin{equation}
    \tau_N(\psi_\mu) = \frac{(|\mu|^4 + 4\Re(\mu^2) + 4)^2}{(2+|\mu|^2)^4}.
\end{equation}
For $\mu = 0$ ($|\mathrm{GHZ}_4\rangle$ state), $\tau_N = 1$, while for $\mu = i \sqrt{2}$ (tetrahedron state), $\tau_N = 0$. For any even $N \geqslant 4$, we have $\tau_N(\mathrm{GHZ}_N) = 1$~\cite{Wong_PRA_63}, while $\tau_N(P_N) = 0$.

\section{Conclusion}
\label{sec:conclusion}

As a conclusion, in this paper we have formalized a general criterion to identify whether a symmetric state is maximally entangled or not in terms of the maximal mixedness of its one-qubit reduced density operator. This criterion is straightforwardly checked if the symmetric states are expressed in the symmetric Dicke state basis. We then have given two physical interpretations of these maximally entangled symmetric (MES) states~: they are the only states for which the expectation value of the associated collective spin vanishes, as well as in corollary the dipole moment of the Husimi function. We have identified that MES states actually coincide with anticoherent spin states to order~1~\cite{Zimba_EJTP_3} as well as with unpolarized light states to order 1~\cite{Luis02,Sanchez-Soto13}. More generally, anticoherent states to order $t>1$ are symmetric states with maximally mixed $t$-qubit reduced density operators in the symmetric subspace (and incidentally maximally mixed $\rho_{t'},\;\forall\;t'\leqslant t$) and are the only states characterized by a Husimi function with vanishing multipolar moments up to order $2^t$~\cite{Giraud}. We have then given a general non-existence criterion of MES states within SLOCC classes. We have shown in particular that the symmetric Dicke state SLOCC classes never contain MES states, with the only exception of the balanced Dicke state class for even numbers of qubits. We have analyzed exhaustively the $4$-qubit case and identified all MES states for this system. These states are the Dicke state $|D_4^{(2)}\rangle$ as well as all states $|\psi_\mu\rangle$ of Eq.~(\ref{nongen4}). Among these states, only the tetrahedron state $|\psi_{\mu=i\sqrt{2}}\rangle$ is also anticoherent to order $2$. We finally have studied the entanglement content of MES states with respect to the geometric and barycentric measures of entanglement, as well as to the generalized $N$-tangle. This entanglement content has been exhaustively analyzed in the $4$ qubit case. We have shown that MES states do not maximize necessarily these entanglement measures, especially when the number of qubits exceeds 4. The geometric measure of entanglement of MES states is ensured to be larger than or equal to $1/2$, the equality being only met, up to local unitaries, for GHZ states (the Bell state for the $2$-qubit system).

\acknowledgments
The authors would like to thank Olivier Giraud for fruitful discussions. T.B. acknowledges the financial support from the Belgian F.R.S.-FNRS through IISN Grant 4.4512.08.

\appendix
\section{Reduced density matrices of symmetric states}\label{app:red_dens_mat}

In this appendix, we provide a compact expression for the $t$-qubit reduced density matrices $\rho_t = \mathrm{tr}_{\neg t} (|\psi_S\rangle \langle \psi_S|)$ ($t = 1, \ldots, N-1$) of any $N$-qubit symmetric states $\ket{\psi_{S}}$.

We start by noting that the symmetric $N$-qubit Dicke states $|D_N^{(k)}\rangle$ can be written as a sum of tensor products of symmetric Dicke states with smaller number of qubits. We have for every $t = 1, \ldots, N-1$
\begin{equation}\label{Dickedecomp}
\ket{\tilde{D}_{N}^{(k)}}=\sum_{q=0}^{t}\ket{\tilde{D}_{t}^{(q)}}\otimes\ket{\tilde{D}_{N-t}^{(k-q)}}
\end{equation}
with
\begin{equation}\label{Dtilde}
\ket{\tilde{D}_{N}^{(k)}} = \sqrt{C_N^k}\,\ket{D_{N}^{(k)}},
\end{equation}
where $C_N^k$ is the binomial coefficient $\left(\begin{smallmatrix}N\\k\end{smallmatrix}
\right)$ with the usual convention $C_N^k = 0$ if $k < 0$ or $k > N$. Any symmetric state $\ket{\psi_S}=\sum_{k=0}^N d_k \ket{D_{N}^{(k)}}$ can thus be written as
\begin{align}
\ket{\psi_{S}} & = \sum_{k=0}^{N}\sum_{q=0}^{t}\ket{\tilde{D}_{t}^{(q)}}\otimes\left(\frac{d_k}{\sqrt{C_{N}^{k}}}\,\ket{\tilde{D}_{N-t}^{(k-q)}}\right)\\
& = \sum_{q=0}^t \ket{{D}_t^{(q)}}\otimes\ket{v_t^{(q)}}, \label{psidecomprho}
\end{align}
where we have introduced the $(N-t)$-qubit states ($t = 1, \ldots, N-1$; $q=0,\ldots, t$)
\begin{equation}
\ket{v^{(q)}_t} = \sqrt{C_t^q}\,\sum_{k=0}^{N-t} \frac{d_{k+q}}{\sqrt{C^{k+q}_{N}}}\,\ket{\tilde{D}_{N-t}^{(k)}}.
\end{equation}
The $t$-qubit reduced density matrices in the Dicke state basis then follow from Eq.~(\ref{psidecomprho}) to correspond to the $(t+1)\times(t+1)$ Gram matrix of the vectors $\ket{v^{(q)}_t}$,
\begin{equation}
\rho_t = \left(\begin{array}{ccc}
\braket{v_{t}^{(0)}}{v_{t}^{(0)}} & \cdots & \braket{v_{t}^{(0)}}{v_{t}^{(t)}}\\
\vdots & \ddots & \vdots \\
\braket{v_{t}^{(t)}}{v_{t}^{(0)}} & \cdots & \braket{v_{t}^{(t)}}{v_{t}^{(t)}}
\end{array}\right)
\end{equation}
with
\begin{equation}\label{eq:vtqvtl}
\braket{v^{(q)}_t}{v^{(\ell)}_t}=\sum_{k=0}^{N-t}d_{k+q}^{*}\, d_{k+\ell}\,\Gamma_k^{q\ell},
\end{equation}
where
\be \label{Gamma}
\Gamma_k^{q\ell}=\frac{1}{C_N^{t}}\sqrt{C^{t-q}_{N-k-q}C_{k+q}^{k}C^{t-\ell}_{N-k-\ell}C_{k+\ell}^{k}}.
\ee

\section{Decomposition of symmetric operators as polynomial in collective spin operators}\label{sec:operator_as_pol_spin_col}

In this appendix, we show that any symmetric operator $\hat{O}$ acting on the $(N+1)$-dimensional symmetric subspace of an $N$-qubit system can always be written as a multivariate polynomial in the collective spin operators $\hat{S}_z$ and $\hat{S}_\pm$ of degree at most $N$ and we present a procedure to determine this polynomial. We start by decomposing an arbitrary symmetric operator $\hat{O}$ onto the symmetric Dicke states basis as
\begin{eqnarray}\label{Odecomp}
\hat{O} &=& \sum_{k=0}^N\sum_{\ell=-k}^{N-k} O_{k+\ell,k}\ketbra{D_{N}^{(k+\ell)}}{D_{N}^{(k)}}
\end{eqnarray}
with $O_{k+\ell,k}=\bra{D_{N}^{(k+\ell)}}\hat{O}\ket{D_{N}^{(k)}}$. Next, we show that any operator of the form $\ketbra{D_N^{(k+\ell)}}{D_N^{(k)}}$ with $\ell\geqslant 0$ appearing in the decomposition (\ref{Odecomp}) of $\hat{O}$ is equal to $\hat{S}_+^\ell\,P_{N-\ell}^{(k)}(\hat{S}_z)$ with $P_{N-\ell}^{(k)}(\hat{S}_z)$ some polynomial in $\hat{S}_z$ of degree at most $N-\ell$. The case of operators of the form $\ketbra{D_N^{(k+\ell)}}{D_N^{(k)}}$ with $\ell<0$ follows directly by Hermitian conjugation. The $N-\ell+1$ operators $ \hat{S}_+^{\ell} \hat{S}_z^m$ for $m=0,\ldots,N-\ell$ form an operator basis for operators whose nonzero entries in the Dicke state basis lie on the $\ell$ diagonal (i.e.,\ whose nonzero entries are $O_{nm}$ with $m=n+\ell$). It follows that any operator $\ketbra{D_N^{(k+\ell)}}{D_N^{(k)}}$ can be written (for $\ell\geqslant 0$) as
\begin{equation}\label{eq:ProjectorOfDiagonal}
\ketbra{D_N^{(k+\ell)}}{D_N^{(k)}} = \hat{S}_+^{\ell} \sum_{m=0}^{N-\ell} \alpha_m^{(k,\ell)} \hat{S}_z^m,
\end{equation}
where the coefficients $\alpha_m^{(k,\ell)}$ obey the linear system of equations
\begin{equation}\label{eqalpham}
\sum_{m=0}^{N-\ell} A_{nm}\, \alpha_m^{(k,\ell)}=\delta_{nk}
\end{equation}
with $n,k=0,\ldots, N-\ell$ and
\be \label{matrixA}
A_{nm}=\sqrt{\prod_{p=n}^{n+\ell-1}(N-p)(p+1)}\:(n-N/2)^m.
\ee
Whenever $\ell=0$, the square root in Eq.~(\ref{matrixA}) should be replaced by $1$. Equation~(\ref{eqalpham}) is obtained by taking the matrix elements of (\ref{eq:ProjectorOfDiagonal}) between the Dicke states $\ket{D_N^{(n+\ell)}}$ and $\ket{D_N^{(n)}}$ for $n=0,\ldots,N-\ell$ and using the fact that Dicke states are simultaneous eigenstates of $\hat{\mathbf{S}}^2$ and $\hat{S}_z$. The matrix $A$ defined by its entries~(\ref{matrixA}) is invertible because it is the product of an invertible diagonal matrix with an invertible Vandermonde matrix with evenly spaced set of ordinates $\{(n-N/2) : n=0,\ldots, N-\ell\}$~\cite{Turner_NTR}. Since $A$ is invertible, the linear system of equations \re{eqalpham} has a unique solution, which yields the desired decomposition \re{eq:ProjectorOfDiagonal}.

\section{Components of the reduced density matrices $\rho_t$ in terms of expectation values of collective spin operators}\label{sec:reduced_density_matrix_spin_coll}

In this appendix, we show that the matrix elements of the $t$-qubit reduced density matrices $\rho_t$ can be written as expectation values of polynomials of degree $t$ in the collective spin operators.

The $\rho_t$ matrix elements $\bra{D_t^ {(q)}}\hat{\rho}_t\ket{D_t^{(\ell)}}$ (see Eq.~\re{eq:vtqvtl}) can be written in the form
\begin{align}\label{eq:red_mat_spin_col}
\bra{D_t^ {(q)}}\hat{\rho}_t\ket{D_t^{(\ell)}} & = \tr(\hat{\rho}_t \ketbra{D_t^{(\ell)}}{D_t^{(q)}}) \\
&=\tr(\hat{\rho}_S\, \hat{O}_t^{q\ell}) = \langle\psi_S|\hat{O}_t^{q\ell}|\psi_S\rangle
\end{align}
with
\begin{equation}\label{Otql}
\hat{O}_t^{q\ell}= \hat{P}_S(\ketbra{D_t^{(\ell)}}{D_t^{(q)}}\otimes\hat{\mathbb{1}}_{N-t})\hat{P}_S
\end{equation}
which follows from the definition of the partial trace~\cite{Nielsen} and where $\hat{P}_S$ is the projector onto the symmetric subspace. Upon using the decomposition of Appendix~\ref{sec:operator_as_pol_spin_col} to the operator $\hat{O}_t^{q\ell}$ given in Eq.~\re{Otql}, this proves that every matrix element of a $t$-qubit reduced density operator can always be expressed as the expectation value of a polynomial of degree at most $N$ in the collective spin operators. Actually, a polynomial of degree $t$ is even sufficient. Indeed, $\hat{O}_t^{q\ell}$ is the symmetrization of the tensorial product of a symmetric operator acting on the $t$-qubit Hilbert space (hence it can be written as a polynomial of degree $t$ in the $t$-qubit collective spin operators) and of the identity in the $(N-t)$-qubit Hilbert space (hence a polynomial of degree zero in the collective spin operators). The conclusion then follows from the definition of the collective spin operators. Indeed, the symmetrization of the tensorial product of two polynomials of degree $r$ and $s$ in collective spin operators acting in the subspaces $\mathcal{H}_t$ and $\mathcal{H}_{N-t}$, respectively, can be written as a polynomial of degree $r+s$ in collective spin operators acting in the global Hilbert space $\mathcal{H}_N\equiv\mathcal{H}_t\otimes \mathcal{H}_{N-t}$.

\section{SLOCC representatives of all
$4$-qubit $\mathcal{D}_{1,1,1,1}$ states}\label{app:slocc4qubit}

In this appendix, we show that any $\mathcal{D}_{1,1,1,1}$-type $4$-qubit symmetric state is SLOCC equivalent to one of the MES states
\begin{equation}\label{eq:4qb}
\ket{\psi_{\mu}} = \frac{1}{\sqrt{2 + |\mu|^2}} (\ket{D_4^{(0)}}+ \mu \ket{D_4^{(2)}}+\ket{D_4^{(4)}}),
\end{equation}
with $\mu$ a $c$-number in the bounded domain
\begin{align}
    \mathit{S} = & \{ \mu \in \mathbb{C} : \Re(\mu) \geqslant 0, \Im(\mu) \geqslant 0, \nonumber \\
    & |\mu - \sqrt{2/3}| \leqslant 2 \sqrt{2/3}, \mu < \sqrt{2/3} \textrm{ if } \Im(\mu) = 0\}.
\end{align}
These states were introduced in~\cite{Bastin-Krins_PREL_103} in the context of the classification of the $4$-qubit symmetric states, but the restricted domain $\mathit{S}$ where distinct $\mu$ define SLOCC inequivalent states was not discussed and too quickly shortcut. In Ref.~\cite{Aul11}, SLOCC representatives in the $\mathcal{D}_{1,1,1,1}$ state family were also identified, but with states not cumulating the property of being MES. Here we show that this is possible with the states (\ref{eq:4qb}).

We first note that the Majorana representation (\ref{mr}) of any symmetric state expressed in the Dicke basis $|\psi_S\rangle = \sum_{k=0}^{N} d_k |D_N^{(k)}\rangle$ is obtained by finding the $M\leqslant N$ roots $z_{m}$ of the polynomial
\begin{equation}
\label{Majpol}
P(z)=\sum_{k=0}^{N}(-1)^{k}\sqrt{C_{N}^{k}}\,d_{k}z^{k}
\end{equation}
and applying the (inverse) stereographic projection from the complex
plane onto the Bloch sphere through the relation $z_{m}=\cot(\theta_{m}/2)e^{-i\varphi_{m}}$, with $(\theta_m, \varphi_m)$ the Bloch sphere coordinates of the Majorana points.
The remaining $N-M$ points are all located at the north pole of the Bloch sphere ($\theta_{m} = 0$)~\cite{Bastin-Krins_PREL_103}. Equivalently, the "roots" $z_{m}$ ($M<m\leqslant N$) can be formally set to $\infty$. For the states $\ket{\psi_{\mu}}$ of Eq.~(\ref{eq:4qb}), setting
\begin{equation}
\mu=-\frac{1}{\sqrt{6}}\left(z^2+\frac{1}{z^2}\right)
\end{equation}
makes the roots $z_m$ ($m = 1, \ldots, 4$) take the simple form $\pm z$ and $\pm 1/z$. They are all distinct as long as $z^2 \neq \pm 1$, i.e., as long as $\mu \neq \pm \sqrt{2/3}$. For these two specific values of $\mu$, the states $|\psi_{\mu}\rangle$ are LU equivalent to the Dicke state $|D_4^{(2)}\rangle$ and are not of the $\mathcal{D}_{1,1,1,1}$ type.

Applying a SLOCC transformation on a symmetric state is equivalent to applying a Moebius transformation (MT) $M(z)$ on its polynomial roots~\cite{Ribeiro_PRL_106, Aul11}. The most general MT reads
\begin{equation}
M(z) = \frac{a z + b}{c z + d},
\end{equation}
where $a,b,c,d\in\mathbb{C}$ and $ad-bc\ne 0$. Moebius transformations form a group, such that the composition of two or more MT is also an MT.

A symmetric local unitary transformation applied on a symmetric state has the effect of a rigid rotation of the corresponding Majorana points on the Bloch sphere. It is thus always possible to take by LU a Majorana point of any symmetric states to the north pole of the Bloch sphere. Doing so for any $\mathcal{D}_{1,1,1,1}$ $4$-qubit symmetric states yields a state with one root of the polynomial $P(z)$ equal to $\infty$, such that the four distinct roots are now given by $z_1,\, z_2,\, z_3,\, \infty$. We then apply a first Moebius transformation,
\begin{equation}
M_1(z_1,z_2,z) = \frac{z-z_1}{z_2-z_1}
\end{equation}
to take $z_1$ to $0$ and $z_2$ to $1$. After this SLOCC transformation, the four roots associated with the state are $\tilde{z}_1=0,\, \tilde{z}_2=1,\, \tilde{z}_3=(z_3-z_1)/(z_2-z_1),$ and $\tilde{z}_4=\infty$. We then apply a second Moebius transformation $M_2(z_0,z)$ in order to take $\tilde{z}_1$ to $z_0$, $\tilde{z}_2$ to $-z_0$, and $\tilde{z}_4$ to $1/z_0$ with $z_0\in\mathbb{C}\setminus\lbrace 0,\pm 1, \pm i \rbrace$. This transformation is explicitly given by
\begin{equation}
M_2(z_0,z) = \frac{ 2 z - z_0(z_0+1/z_0)}{2 z_0 z - (z_0 + 1/z_0)}.
\end{equation}
In order to put the state to the desired form (\ref{eq:4qb}), it suffices to choose $z_0$ such that
\begin{equation}
M_2(z_0,\tilde{z}_3) = -1/z_0.
\end{equation}
This is the case if $z_0$ is a square root of $2 \tilde{z}_3 - 1 + 2 \delta$,
with $\delta$ a square root of $\tilde{z}_3(\tilde{z}_3 - 1)$.
In summary, the composition $M_2(z_0(\tilde{z}_3),z)\circ M_1(z_1,z_2,z)$ is a state-dependent Moebius transformation that can be applied to any $\mathcal{D}_{1,1,1,1}$-type states in order to put it to the form (\ref{eq:4qb}), after an LU has been applied to take one point at the north pole of the Bloch sphere. At this stage, the obtained state $|\psi_{\mu}\rangle$ is not yet ensured to be such that $\mu \in \mathit{S}$. We show hereafter how to get to this final step.

 All states $|\psi_{\mu}\rangle$ of Eq.~(\ref{eq:4qb}) with arbitrary $\mu \in \mathbb{C}$ are not SLOCC inequivalent to each other. This is only the case for $\mu$ in the restricted domain $\mathit{S}$ and all states with $\mu$ out of this domain are LU equivalent to one of these states with $\mu \in \mathit{S}$. To prove this, we first note that if two arbitrary states $|\psi_{\mu}\rangle$ and $|\psi_{\mu'}\rangle$ are SLOCC equivalent, then they are also necessarily LU equivalent. This is because MES states are unique up to local unitaries within their SLOCC classes~\cite{Verstraete_PRA_68} and so are all states $|\psi_{\mu}\rangle$ (see Sec.~\ref{sec:On-the-uniqueness}). LU-equivalent symmetric states can be transformed into each other using an identical local unitary for each qubit~\cite{Mar11}. Considering the most general expression of single-qubit unitary operations $U$, one can identify all local unitaries that transform a state $|\psi_{\mu}\rangle$ into a state of the same type, $|\psi_{\mu'}\rangle$, i.e., that implement in the complex plane $\mu \rightarrow \mu'(\mu)$ transformations. Up to a global phase and to the identity operation, these LUs are exhaustively given by the Pauli matrices $\sigma_x$, $\sigma_y$ and $\sigma_z$ ($\mu' = \mu$)~\cite{footnote3}, by the matrices
\begin{equation}
    U_1 = \left( \begin{array}{cc} 1 & 0 \\ 0 & i \end{array} \right), U_2 = \frac{1}{\sqrt{2}}\left( \begin{array}{cc} 1 & 1 \\ 1 & -1 \end{array} \right), U_3 = \frac{1}{\sqrt{2}}\left( \begin{array}{cc} 1 & i \\ i & 1 \end{array} \right),
\end{equation}
and by any composition of these unitaries. The symmetric LUs $U_1^{\otimes 4}$, $U_2^{\otimes 4}$, and $U_3^{\otimes 4}$ implement the Moebius transformations $\mu' = -\mu$, $\mu' = 2 (\sqrt{6} - \mu)/(\sqrt{6}\mu + 2)$, and $\mu' = 2 (\sqrt{6} + \mu)/(\sqrt{6}\mu - 2)$, respectively. Each of these transformation coincides with its inverse and maps the upper part of the complex plane to the lower part and vice versa ($\Im(\mu')$ and $\Im(\mu)$ have opposite signs). For $U_2^{\otimes 4}$, the right part of the complex plane ($\Re(\mu) \geqslant 0$) is mapped into the closed disk of radius $2\sqrt{2/3}$ and centered on $\sqrt{2/3}$ (the single point $-\sqrt{2/3}$ excluded), while the right-upper part of the disk (where $\Re(\mu)$ and $\Im(\mu) \geqslant 0$) is mapped into its right-lower part and vice versa. It also maps region II of Fig.~\ref{fig:EG} to the complex conjugate of region I and vice versa. For real $\mu$, the interval $]\sqrt{2/3}, \sqrt{6}]$ is mapped into $[0,\sqrt{2/3}[$ and vice versa.  For $U_3^{\otimes 4}$ and real $\mu$, the interval $[0, \sqrt{2/3}[$ is mapped into $]-\infty,-\sqrt{6}]$. As a consequence of all this, a right sequence of the local unitaries $U_1^{\otimes 4}$ and $U_2^{\otimes 4}$ applied alternatively and at most twice on any states $|\psi_{\mu}\rangle$ with $\mu \notin \mathit{S}$ transform the state into an LU equivalent state $|\psi_{\mu'}\rangle$ with $\mu' \in \mathit{S}$. We are finally ensured that all states with $\mu \in \mathit{S}$ are SLOCC inequivalent; otherwise they would be LU-equivalent and this is impossible since all LUs connecting $|\psi_{\mu}\rangle$-type states together are exhaustively listed here above and none of them keeps $\mu$ inside the domain $\mathit{S}$.


\begin{thebibliography}{10}
\bibitem{footnote3} As a corollary, this means that all states $|\psi_{\mu}\rangle$ are invariant under the action of the same Pauli matrices over each qubit~: $|\psi_{\mu}\rangle = \sigma_i^{\otimes 4} |\psi_{\mu}\rangle$, for $i=x,y,z$ and $\forall \mu \in \mathbb{C}$. Up to a global phase and to the identity operation, these transformations are the only LUs that leave invariant all states $|\psi_{\mu}\rangle$.
\end{thebibliography}
\end{document}